\documentclass[12pt]{article}
\usepackage{graphicx}
\usepackage{amsmath}
\usepackage{amssymb}
\usepackage{caption2}
\begin{document}
\bibliographystyle{prsty}
\begin{center}
{\large {\bf \sc{   Decay width of the pentaquark state $\Theta^+(1540)$ with QCD sum rules }}} \\[2mm]
Zhi-Gang Wang$^{1}$ \footnote{Corresponding author; E-mail,wangzgyiti@yahoo.com.cn.  }, Wei-Min Yang$^{2}$ and Shao-Long Wan$^{2} $    \\
$^{1}$ Department of Physics, North China Electric Power University, Baoding 071003, P. R. China \\
$^{2}$ Department of Modern Physics, University of Science and Technology of China, Hefei 230026, P. R. China \\
\end{center}

\begin{abstract}
In this article, we take the point of view that the pentaquark state
$\Theta^+(1540)$ has negative parity, and choose the
diquark-triquark type interpolating current to calculate the strong
coupling constant $g_{\Theta NK}$ in the QCD sum rules  approach.
Our numerical results  indicate  the values of the strong coupling
constant $g_{\Theta NK}$ are very small, $|g_{\Theta
NK}|=0.175\pm0.084$, and the  width $\Gamma_\Theta <4MeV$,
  which can explain
the narrow width $\Gamma \leq 10 MeV$ naturally.
\end{abstract}


\section{Introduction}

Several collaborations have reported the observation of the new
baryon $\Theta^+(1540)$ with positive strangeness and minimal
quark contents  $udud\bar{s}$ \cite{exp2003}. The existence of
such an exotic state with narrow width $\Gamma < 15 MeV$ and
$J^P={\frac{1}{2}}^+$ was first predicted in the chiral quark
soliton model, where the $\Theta^+(1540)$ is a member of the
baryon antidecuplet $\overline{10}$ \cite{Diakonov97}. The
discovery has opened a new field of strong interactions   and
provides a new opportunity for a deeper understanding of the low
energy QCD. Intense theoretical investigations have been motivated
 to clarify the quantum numbers and to understand the
under-structures of the pentaquark state $\Theta^+(1540)$
\cite{ReviewPenta}. The zero of the third component of isospin
$I_3=0$ and the absence of isospin partners suggest  that the
baryon $\Theta^+(1540)$ is an isosinglet, while the spin and
parity have not been experimentally determined yet and  no
consensus has ever been reached  on the theoretical side. The
extremely narrow width  below $10MeV$ puts forward a serious
challenge to all theoretical models, in the conventional
uncorrelated quark models  the expected width is of the order of
several hundred $MeV$, since the strong decay $\Theta^{+} \,
\rightarrow K^+ N$ is Okubo-Zweig-Iizuka (OZI) super-allowed.

 In this article, we take the point of view that the quantum
numbers of the pentaquark state  $\Theta^{+}(1540)$ are
$J^P={\frac{1}{2}}^-$ , $I=0$ , $S=+1$, and study its decay width
within the framework of  the QCD sum rules approach
\cite{Shifman79,Narison04,Nielsen05}.

 The article is arranged as follows:   we derive the QCD sum rules
  for the strong coupling constant of the
pentaquark state $\Theta^+(1540)$ $g_{\Theta N K}$ in section II; in
section III, numerical results; section IV is reserved for
conclusion.

\section{QCD sum rules for the coupling constant $g_{\Theta N K}$}
In  the following, we write down the three-point correlation
function \cite{Nielsen05,Zhu03},
\begin{eqnarray}
 \Gamma(p,q) &=& \int d^4x  d^4y    \, e^{ip \cdot
x} e^{-iq\cdot y}  \langle 0|T\{\eta _N(x)j_{K}(y)
\bar{\eta}_{\Theta}(0)\}|0 \rangle\,,
\end{eqnarray}
where the $\eta_N$, $j_{K}$ and
$\overline{\eta}_{\Theta}=\eta_{\Theta}^+\gamma^0$ are the
interpolating currents for the neutron, K meson and pentaquark state
$\Theta^+(1540)$ respectively,
\begin{eqnarray}
\eta_{\Theta}(0)&=&{1\over \sqrt{2}} \epsilon^{abc} \left\{u^T_a(0)
C\gamma_5 d_b (0)\right\} \{ u_e (0) {\bar s}_e (0) i\gamma_5 d_c(0)
- d_e (0) {\bar s}_e (0) i\gamma_5 u_c(0)  \} \, , \\
j_{K}(y)&=&\bar{s}(y)i\gamma_{5}u(y)\, , \\
\eta_{N}(x)&=&\epsilon^{abc} ({d}^{T}_a(x)C\gamma _{\mu}d_{b}(x))
\gamma _5 \gamma ^{\mu}u_c(x)\, .
\end{eqnarray}
For the pentaquark state $\Theta^+(1540)$, we use the
diquark-triquark type interpolating current which can give
satisfactory mass  and stable magnetic moment \cite{Zhu03,ZhuWang}.
The pseudoscalar mesons $\pi$ and $K$ can be taken as both Goldstone
bosons and quark-antiquark bound states, we can use the partial
conservation  of axial current (PCAC) in constructing the
interpolating currents,
$\partial_\mu(\bar{s}(x)\gamma^\mu\gamma_{5}u(x))$,
\begin{eqnarray}
\partial_\mu(\bar{s}(x)\gamma^\mu\gamma_{5}u(x))&=&(m_s+m_u)\bar{s}(x)i\gamma_{5}u(x),\nonumber \\
\langle0|\partial_\mu(\bar{s}(0)\gamma^\mu\gamma_{5}u(0))|K(q)\rangle &=&(m_s+m_u)\langle0|\bar{s}(0)i\gamma_{5}u(0)|K(q)\rangle,\nonumber\\
&=& f_{K}q^2 = f_{K}m_K^2 \nonumber.
\end{eqnarray}
If we take the $\bar{s}(x)\gamma^\mu\gamma_{5}u(x)$ as the
interpolating current, more care has to be taken about the possible
contaminations from the axial-vector mesons, furthermore, the
calculation will be more tedious ( with
$\partial_\mu(\bar{s}(x)\gamma^\mu\gamma_{5}u(x))$ ). The matrix
element of the pseudoscalar current between the vacuum and $K$ state
can be taken as
\begin{eqnarray}
\langle0|\bar{s}(0)i\gamma_{5}u(0)|K(q)\rangle=\lambda_K=\frac{f_K
m_K^2}{m_u+m_s},
\end{eqnarray}
the values of the $\lambda_K$ depend on the masses of the $s$ and
$u$ quarks which have uncertainties, our numerical results indicate
small variations of those masses will not lead to large changes
about the values of the coupling constant  $g_{\Theta NK}$. For the
neutron, we take the Ioffe current \cite{Ioffe}. The Fierz
re-ordering of the interpolating currents $\eta_{\Theta}$ and
$\eta_N$ can lead to the following sub-structures,
\begin{eqnarray}
\eta_N &=&  \epsilon^{abc} \left\{ ( u_a^T C d_b ) \gamma_5 d_c - (
u_a^T C \gamma_5 d_b )
d_c \right\}, \\
\eta_{\Theta} &=&{1
\over4\sqrt{2}}\epsilon^{abc}(u^{T}_aC\gamma_5d_b)\left\{-d_c(\bar s
i\gamma_5 u)
  -\gamma^\mu d_c(\bar s i\gamma_5\gamma_\mu u)\right.\cr
  &&\left.-\frac{1}{2}\sigma^{\mu\nu}d_c(\bar s i\sigma_{\mu\nu}\gamma_5 u)
  -\gamma^\mu\gamma_5 d_c(\bar s i\gamma_\mu u)
  -\gamma_5 d_c(\bar s iu)-(u \leftrightarrow d)\right\}.
\end{eqnarray}
A naive result  of the Fierz re-ordering may be the appearance of
the reducible contributions with the sub-structure of $udd-u\bar{s}$
 (i.e. $N-K$) clusters   in the two-point correlation function
\cite{Reducible},
\begin{eqnarray}
 \Pi(p) &=& \int d^4x   \, e^{ip \cdot
x}   \langle 0|T\{\eta _\Theta(x) \bar{\eta}_{\Theta}(0)\}|0
\rangle\,,
\end{eqnarray}
however, in our calculations with the interpolating current
$\eta_{\Theta}$ in Eq.(2), no such factorable  $udd-u\bar{s}$ terms
appear, so there are no reducible $N-K$ contributions to the
correlation function $\Pi(p)$.  The re-ordering in the Dirac spin
space is always  accompanied  with  the color re-arrangement, which
involves the underlying dynamics.   If we want to factorize out some
$N-K$ contributions from the $\Pi(p)$ ( with the current
$\eta_{\Theta}$ in Eq.(2) ), tedious manipulations about the
re-ordering in the color and Dirac spin space must be done. There
are no direct $N-K$ ( or $udd-u\bar{s}$ ) components in the
interpolating current $\eta_{\Theta}$,  which can readily  decay to
the $NK$ final state and result in large width. If the
$\Theta^+(1540)$ is really a pentaquark state not a $N-K$ molecule,
as the $\Theta^+(1540)$ lies above the $NK$ threshold and no need
for additional quark-antiquark pairs  creation in decay, the decay
must be OZI super-allowed and the width is supposed   be large, say,
about several hundred $MeV$; to produce the narrow width, some huge
energy barriers are needed to stabilize the $\Theta^+(1540)$ in case
the kinematical interpretation can not work  here. The appearance of
the $N-K$ component in the Fierz re-ordering maybe manifest the
possibility ( not the probability ) of the evolution from the
$\Theta^+(1540)$ to the $NK$ final state without net quark-antiquark
pairs  creation ( maybe the quark-antiquark pairs created and
annihilated subsequently ), which is significantly in contrast to
the conventional baryons, however, we have no knowledge about the
detailed process of the evolution. The $\Theta^+(1540)$ may evolve
to the $N-K$ final state, and the $N-K$ final  state  is not
presented in the components of the initial pentaquark state
$\Theta^+(1540)$, how to implement the evolution with  small
probability involves complex quark-gluon interactions,  whether just
re-arrangement in the color space, or creation and annihilation of
quark-antiquark pairs. In additive constituent quark models, whether
or not additional relative P wave is introduced to changed the
ground states from negative parity to positivity parity, special
configurations are needed to take into account the narrow decay
width  as  results of small overlaps  of the internal and external
$N-K$ wave-functions \cite{WaveAdd}, the kinematical interpretations
based on the color-flavor-spin ( i.e. $SU(3)_c\times SU(3)_f\times
SU(2)_s$ ) group theory resort  to the possible small overlaps,
 how to realized the small overlaps needs complex
re-ordering in the color-flavor-spin space, if there are really no
energy barriers to prevent the re-arrangement.  While in the cluster
quark models, typically, the diquark-diquark-antiquark model
\cite{JW03} and diquark-triquark model \cite{KL}, extra barriers,
for example, relative P waves, are introduced dynamically to prevent
the ready  decay. In fact, the re-ordering in the color and Dirac
spin space involves  complex strong interactions, and  we know
little now about the dynamics which determine the under-structures
of the exotic pentaquark states. The mismatches between the
color-flavor-spin states in the initial pentaquark and final
baryon-meson color singlet can result in suppression  of the decay
naturally \cite{WaveAdd,JM04}. Due to the spontaneous  breaking of
the chiral symmetry and the Goldstone nature of the pseudoscalar
mesons $\pi$, $K$ and $\eta$, the quarks may have direct
interactions with the pseudoscalar mesons, which lead to the success
of some chiral quark models. The dominating interactions which
determine the exotic pentaquark states be color-spin type
$\lambda^c_i \cdot\lambda^c_j \sigma^i \cdot\sigma^j $ or
flavor-spin type $\lambda^f_i\cdot \lambda^f_j \sigma^i
\cdot\sigma^j $ are still in hot debates \cite{color-spin-flavor},
the naive Fierz re-ordering can lead to direct $N-K$ component in
the $\Theta^+(1540)$ will not work here. If there are really some
$N-K$ components in the interpolating current $\eta_\Theta$, they
should be factorized out, the remainder can not have the correct
quantum numbers to interpolate  the $\Theta^+(1540)$. In the QCD sum
rules, we construct the interpolating currents with the same quantum
number as the corresponding mesons and baryons, that is enough; the
knowledge about the structures of the hadrons can be of much help in
the constructing.

In Ref.\cite{Narison04}, the narrow decay width is attributed to the
minor breaking of chirality conservation,   the color re-arrangement
due to the hard gluon-exchange can result in strong suppression of
the decay,  $\Gamma_{\Theta}  \sim \alpha^2_s \, <0|\overline{q}
q|0>^2$. In the article, we take quantitative analysis of the decay
width within the framework of the QCD sum rules approach.

 The diquark-triquark type interpolating current $\eta_\Theta(x)$ is more likely
related to a negative parity pentaquark state, in this work, we make
 assumption that the parity of the $\Theta^+(1540)$ to be negative and study the
decay width with the  following  Lagrangian density,
\begin{eqnarray}
{\cal L}&=&ig_{\Theta NK}\bar{\Theta}KN  .\ \ \
\end{eqnarray}
 According to the
basic assumption of current-hadron duality in the QCD sum rules
approach \cite{Shifman79}, we insert a complete series of
intermediate states satisfying the unitarity principle with the same
quantum numbers as the interpolating currents $\eta_N$, $j_{K}$ and
$\overline{\eta}_{\Theta}$
 into the correlation function in
Eq.(1)  to obtain the hadronic representation. After isolating the
double-pole and single-pole terms of the lowest ground  states, we
get the following result \cite{Nielsen05,Ioffe84},
\begin{eqnarray}
 \Gamma(p,q)&=&\left\{ \frac{ g_{\Theta NK}
\lambda_{\Theta}\lambda_{N}\lambda_{K}} { m_{\Theta}^2-
p'^2}\frac{1}{(m_{N}^2- p^2)(m_{K}^2- q^2)}
+\left[\frac{A(p'^2,q^2)}{ m_{N}^2-p^2}+ \frac{B(p'^2,p^2)}{m_{K}^2-
q^2}
\right]+\cdots \right\} \nonumber\\
&&  \left \{\sigma^{\mu\nu} q_{\mu}p_{\nu} \,\, +
\cdots\right\}+\cdots\, , \nonumber \\
&=&\left\{ \frac{ g_{\Theta NK}
\lambda_{\Theta}\lambda_{N}\lambda_{K}} { m_{\Theta}^2-
p'^2}\int_0^{\infty}ds \int_0^{\infty}dt
\frac{\delta(s-m_{N}^2)\delta(t-m_{K}^2 )}{(s- p^2)(t- q^2)} +
\right.\nonumber\\
&&\left. \left[\int_0^\infty ds \frac{A(p'^2,q^2)\delta(s-m_{N}^2)}{
s-p^2}+ \int_0^\infty dt \frac{B(p'^2,p^2)\delta(t-m_{K}^2)}{t- q^2}
\right]+\cdots \right\} \nonumber\\
&&  \left \{\sigma^{\mu\nu} q_{\mu}p_{\nu} \,\, +
\cdots\right\}+\cdots\, ,
\end{eqnarray}
with
\begin{eqnarray}
A(p'^2,q^2)&=& \int_{m^2_{K^*}}^\infty dt
\frac{\rho_A(p'^2,t)}{t-q^2},
\\
B(p'^2,p^2)&=& \int_{m^2_{N^*}}^\infty ds
\frac{\rho_B(p'^2,s)}{s-p^2},
\end{eqnarray}
where the following definitions have been used,
\begin{eqnarray}
\langle 0|\eta _N|N(p,s)\rangle &=& \lambda_{N} u(p,s) \,,\nonumber\\
\langle \Theta(p',s')|\bar{\eta}_{\Theta}|0 \rangle &=&
\lambda_{\Theta}\bar{u}(p',s')\, . \nonumber
\end{eqnarray}
The coupling constants $\lambda_{N}$ and  $\lambda_{\Theta}$ can be
determined from the two-point QCD sum rules, for the
$\lambda_{\Theta}$, we use the correlation function $\Pi(p)$,
substitute the $\eta_N$ for the $\eta_\Theta$ in Eq.(8), we can
obtain the $\lambda_{N}$. In this article, we choose the Dirac
tensor structure $\sigma^{\mu\nu} q_{\mu}p_{\nu}$ for analysis,
while in Ref.\cite{Nielsen05}, the authors take the structure
$\gamma_5 \sigma^{\mu\nu} q_{\mu}p_{\nu}$ . Here the residues of the
single-pole terms $A(p'^2,q^2)$ and $B(p'^2,p^2)$ have complex
dependence on the transitions between the ground states and high
resonances ( or continuum states ). We have no knowledge about  the
transitions, even the existence of the $\Theta^+(1540)$ is not
firmly established. However, the contributions from the
pole-continuum transitions
 are not exponentially suppressed compared with the double-pole terms, even
after double Borel transform, furthermore, the contributions can be
as large as or larger than the double-pole terms and must be
explicitly included in the sum rules. We only have the fact that the
$\Theta^+(1540)$ lies a little above the $NK$ threshold, the
contributions from the $\Theta^+(1540)$ can be factorized  out,  so
the spectral densities $\rho_A$ and $\rho_B$ can be parameterized as
\begin{eqnarray}
\rho_A(p'^2,t)&=&\frac{EE(p'^2,t)}{m_{\Theta^+}^2-p'^2}, \\
\rho_B(p'^2,s)&=&\frac{FF(p'^2,s)}{m_{\Theta^+}^2-p'^2}.
\end{eqnarray}
The  two unknown functions $EE$ and $FF$ have  complex dependence on
the  transitions between the ground states and high resonances ( or
continuum states ).  From the Eqs.(10-14), we can obtain
\begin{eqnarray}
 \Gamma(p,q)&=&\left\{ \frac{ g_{\Theta NK}
\lambda_{\Theta}\lambda_{N}\lambda_{K}} { m_{\Theta}^2-
p'^2}\frac{1}{(m_{N}^2- p^2)(m_{K}^2- q^2)} +\right.\nonumber\\
&&\left.\frac{ 1} { m_{\Theta}^2-p'^2}\left[\frac{1}{
m_{N}^2-p^2}\int_{m_{K^*}^2}^\infty dt \frac{EE(p'^2,t)}{t-q^2}+
\frac{1}{ m_{K}^2-q^2}\int_{m_{N^*}^2}^\infty
ds\frac{FF(p'^2,s)}{s-p^2}
\right]+\cdots \right\} \nonumber\\
&&  \left \{\sigma^{\mu\nu} q_{\mu}p_{\nu} \,\, +
\cdots\right\}+\cdots\, , \\
&=&\left\{ \frac{ g_{\Theta NK}
\lambda_{\Theta}\lambda_{N}\lambda_{K}} { m_{\Theta}^2- p'^2
}\frac{1}{(m_{N}^2- p^2)(m_{K}^2- q^2)} +\frac{ 1} {
m_{\Theta}^2-p'^2}\left[\frac{CC}{ m_{N}^2-p^2}+ \frac{DD}{
m_{K}^2-q^2}
\right]+\cdots \right\} \nonumber\\
&&  \left \{\sigma^{\mu\nu} q_{\mu}p_{\nu} \,\, +
\cdots\right\}+\cdots\, ,
\end{eqnarray}
here we introduce two constants $CC$ and $DD$ for convenience,
\begin{eqnarray}
CC&=&\int_{m_{K^*}^2}^\infty dt \frac{EE(p'^2,t)}{t-q^2},\\
DD&=&\int_{m_{N^*}^2}^\infty ds\frac{FF(p'^2,s)}{s-p^2}.
\end{eqnarray}
Taking the $CC$ and $DD$ as some unknown constants has smeared the
complex dependence on the energy and high resonance masses ( or
continuum states ), which will certainly impair the prediction
power.
  We have no knowledge about the  transitions between the pentaquark state $\Theta^+(1540)$
 and the excited states ( or high resonances, or continuum states ), the $CC$ and $DD$ can be taken as free parameters
, we  choose the suitable values for the $CC$ and $DD$ to
  eliminate the contaminations from  the single-pole terms to obtain the reliable  sum rules. The contributions from
  the single-pole terms may as large as or larger than the double-pole term, in practical manipulations,
   the $CC$ and $DD$ can be fitted to give stable sum rules with respect to
 variations  of the Borel parameter $M^2$ in a suitable interval. If the final numerical results are insensitive to the
 threshold parameters $s_0$, $t_0$ and there really exists a platform with the variations
 of the Borel parameters $M^2_1$ and $M_2^2$, the
 predictions  make sense.

  The  calculation of  operator product expansion in the  deep Euclidean space-time region is
  straightforward and tedious, here technical details are neglected for simplicity,
  once  the analytical  results are obtained,
  then we can express the correlation functions at the level of quark-gluon
degrees of freedom into the following form through dispersion
relation,
  \begin{eqnarray}
 \Gamma(p,q)&=& \sqrt{2}\left\{\frac{21m_s}{2^{12}4!\pi^6}\int_{0}^{s_0}ds \int_{0}^{t_0}dt
  \frac{ s^2}{s-p^2}\frac{ 1}{t-q^2}
  -\frac{11m_s \langle \bar{q}q\rangle^2}{2^7 3\pi^4p^2}\int_0^{t_0}dt \frac{1}{t-q^2}\right.\nonumber\\
  && \left.+\frac{7\left[\langle \bar{q}q\rangle+\langle \bar{s}s\rangle \right]}{2^9 4! \pi^6 q^2}\int_0^{s_0}ds
  \frac{s^2}{s-p^2}-\frac{11\left[ \langle \bar{q}q\rangle^3+\langle\bar{q}q\rangle^2\langle\bar{s}s\rangle \right]}
  {2^4 3^2  \pi^4p^2q^2}\right\} \sigma^{\mu\nu}p_{\mu}q_\nu+\cdots\, ,
  \end{eqnarray}
We choose $p^2=p'^2=-P^2$ and $q^2=-Q^2$, then take double Borel
transform with respect to the variables $P^2$ and $Q^2$
respectively, match Eq.(16) with Eq.(19), finally we obtain the sum
rules for the strong coupling constant $g_{\Theta NK}$,
\begin{eqnarray}
&&g_{\Theta NK}\lambda_K \lambda_N \lambda_\Theta
e^{-\frac{m_K^2}{M_2^2}}
\frac{e^{-\frac{m_\Theta^2}{M_1^2}}-e^{-\frac{m_N^2}{M_1^2}}}{m_\Theta^2-m_N^2}+Ce^{-\frac{m_K^2}{M_2^2}}
e^{-\frac{m_\Theta^2}{M_1^2}}= \nonumber \\
&&\sqrt{2}\left\{\frac{21m_s M_1^6 M_2^2
E_2(s)E_0(t)}{2^{11}4!\pi^6}+
  \frac{11m_s \langle \bar{q}q\rangle^2 M_2^2E_0(t)}{2^7 3\pi^4}\right.\nonumber\\
  && \left.-\frac{7\left[\langle \bar{q}q\rangle+\langle \bar{s}s\rangle \right]M_1^6E_2(s)}{2^8 4! \pi^6 }
  -\frac{11\left[ \langle \bar{q}q\rangle^3+\langle\bar{q}q\rangle^2\langle\bar{s}s\rangle \right]}{2^4 3 ^2\pi^4}
  \right\}\, ,
\end{eqnarray}
where
\begin{eqnarray}
E_n(s)&=&1-e^{-\frac{s_0}{M_1^2}}
\sum_{k=0}^{n}\left(\frac{s_0}{M_1^2}\right)^k\frac{1}{k!}\, ,
\nonumber \\
E_n(t)&=&1-e^{-\frac{t_0}{M_2^2}}\sum_{k=0}^{n}\left(\frac{t_0}{M_2^2}\right)^k\frac{1}{k!}\,
. \nonumber
\end{eqnarray}
 Here the $C$ ( proportional to the $DD$, as the $CC$ terms are eliminated  ) denotes the contributions from
 the transitions between the ground  and excited states ( or high resonances, or continuum states ),
  we can choose the suitable values for $C$ to
  eliminate the contaminations  to obtain the stable sum rules with the variations of the Borel parameters $M_1^2$ and $M_2^2$.

\section{Numerical Results}
The parameters for the condensates are chosen to be the standard
values \cite{Shifman79},  $\langle \bar{s}s \rangle=(0.8\pm0.1)
\langle \bar{u}u \rangle$, $\langle \bar{q}q \rangle=\langle
\bar{u}u \rangle=\langle \bar{d}d \rangle=-(240\pm 10 MeV)^3$,
 $m_u=m_d=0$ and $m_s=(140\pm10) MeV$. Small variations of those condensates will not
 lead to large  changes about   the numerical
 values.  The coupling constants are taken as $\lambda_N=(2.4\pm0.2)\times 10^{-2} GeV^3$ \cite{Ioffe,Ioffe84,Ioffe82} and
 $\lambda_\Theta=( 1.4 \pm 0.2)\times 10^{-4} GeV^6$ \cite{Zhu03} from the two-point QCD sum rules, see Eq.(8) for example.
 The threshold parameters $s_0$ and $t_0$ are chosen to  vary between $(1.8-2.0) GeV^2$ and
  $(0.8-1.0) GeV^2$ respectively  to avoid possible contaminations from
  high  resonances and continuum states. The Borel parameters are
  taken as $M_2^2=(1.0-1.5)GeV^2$ and $M_1^2=(1.0-2.0)GeV^2$  to obtain the stable sum
  rules. Finally we obtain the values for the coupling constant $|g_{\Theta NK}|$,
 \begin{eqnarray}
 |g_{\Theta NK}|=0.175\pm0.084,
  \end{eqnarray}
  \begin{eqnarray}
 \Gamma_{\Theta}&=& \frac{1}{8\pi m_{\Theta}^3}g_{\Theta
nK}^2 [(m_N + m_{\Theta})^2-m_K^2]
\sqrt{\lambda(m_{\Theta}^2,m_{N}^2,m_K^2)}\,, \nonumber\\
&<&4MeV,  \\
\lambda(m_{\Theta}^2,m_{N}^2,m_K^2)&=& (m_{\Theta}^2+ m_{N}^2
-m_K^2)^2-4m_{\Theta}^2m_{N}^2\,. \nonumber
\end{eqnarray}
which can explain the narrow width $\Gamma \leq 10 MeV$ naturally.
The values of the coupling constant $g_{\Theta NK}$ with the
variations of the threshold parameters ($s_0$, $t_0$) and Borel
parameters ($M_1^2$ , $M_2^2$) are shown in Fig.1 and Fig.2 for
$\langle \bar{s}s \rangle=0.8 \langle \bar{u}u \rangle$, $\langle
\bar{q}q \rangle=-(240MeV)^3$,  $m_s=140 MeV$.

\begin{figure}
 \centering
 \includegraphics[totalheight=7cm]{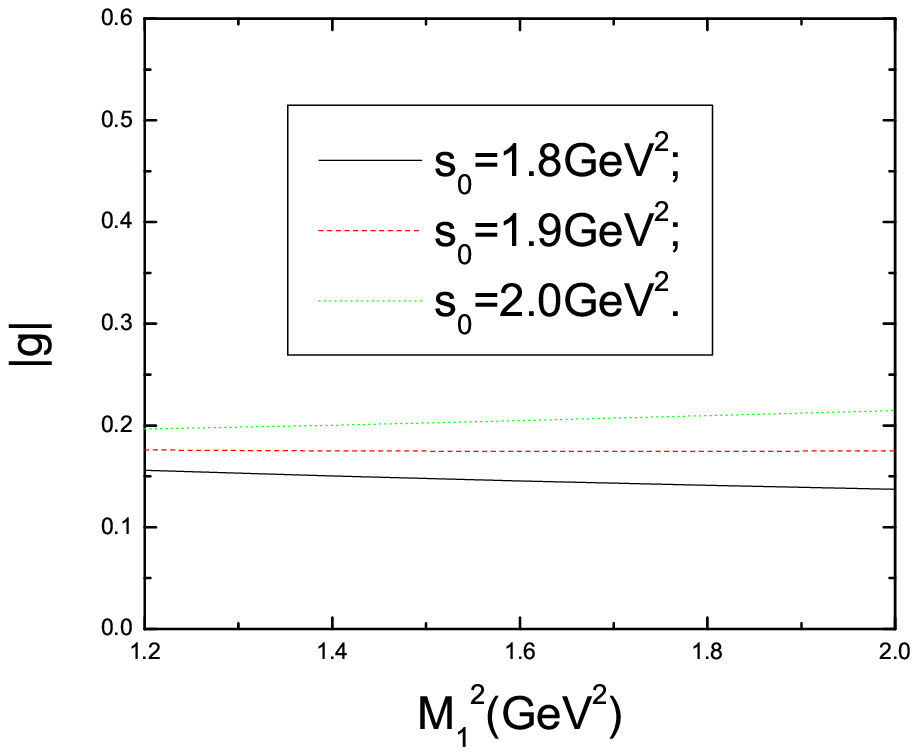}
 \caption{$|g_{\Theta NK}|$ with $M_1^2$ for $M_2^2=1.3GeV^2$, $t_0=0.9GeV^2$. }
\end{figure}

\begin{figure}
 \centering
 \includegraphics[totalheight=7cm]{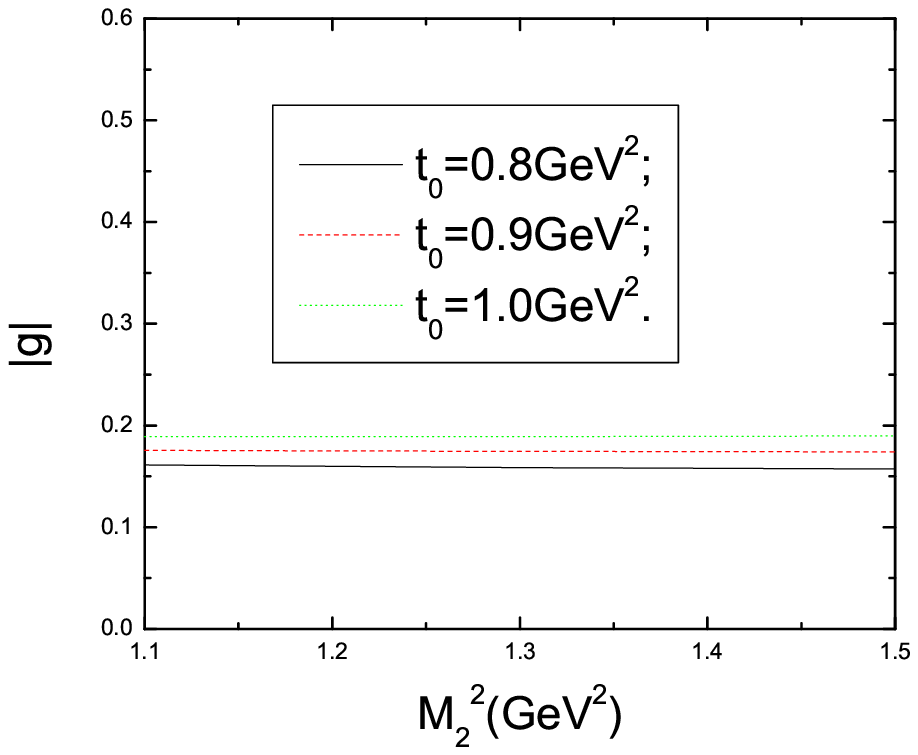}
 \caption{$|g_{\Theta NK}|$ with $M_2^2$ for $M_1^2=1.7GeV^2$, $s_0=1.9GeV^2$}
\end{figure}
\section{Conclusion }
In this article, we take the point of view that the pentaquark state
$\Theta^+(1540)$ has  negative parity, and choose the
diquark-triquark type interpolating current to calculate the strong
coupling constant $g_{\Theta NK}$ within the framework of the QCD
sum rules  approach. Our numerical results  indicate that values of
the strong coupling constant $g_{\Theta NK}$ are very small,
$|g_{\Theta NK}|=0.175\pm0.084$, and the  width $\Gamma_\Theta
<4MeV$,
  which can explain
the narrow width $\Gamma \leq 10 MeV$ naturally.

\section*{Acknowledgment}
This  work is supported by National Natural Science Foundation,
Grant Number 10405009,  and Key Program Foundation of NCEPU. The
authors are indebted to Dr. J.He (IHEP), Dr. X.B.Huang (PKU) and
Dr. L.Li (GSCAS) for numerous help, without them, the work would
not be finished.


\begin{thebibliography}{99}
\bibitem{exp2003}
LEPS collaboration, T. Nakano {\it et al} , Phys. Rev. Lett. {\bf
91} (2003) 012002; S. Kabana, hep-ex/0503020; K. Hicks,
hep-ex/0412048;  and references therein.

\bibitem{Diakonov97}
 D. Diakonov, V. Petrov and M. V. Polyakov,
 Z. Phys. {\bf A359} (1997) 305.

\bibitem{ReviewPenta} M. Oka,   Prog. Theor. Phys. {\bf 112} (2004) 1;
S. L. Zhu,   Int. J. Mod. Phys. {\bf A19} (2004) 3439; S. L. Zhu,
hep-ph/0410002; F. E. Close, hep-ph/0311087;
 and references
therein.


\bibitem{Shifman79} M. A. Shifman, A. I. Vainshtein  and V. I. Zakharov,
 Nucl. Phys. {\bf B147 } (1979) 385, 448; L. J. Reinders, H. Rubinstein and S. Yazaki,
Phys. Rept.  {\bf 127} (1985) 1; S. Narison, World Sci.\ Lect.\
Notes Phys.  {\bf 26} (1989) 1.

\bibitem{Narison04} R. D. Matheus and S. Narison, hep-ph/0412063;
 B. L. Ioffe and A. G. Oganesian, JETP Lett.  {\bf 80} (2004) 386;
A.  G. Oganesian, hep-ph/0410335.


\bibitem{Nielsen05} M. Eidemuller, F. S. Navarra,
M. Nielsen and R. Rodrigues da Silva, hep-ph/0503193.

\bibitem{Zhu03}
S. L. Zhu, Phys. Rev. Lett. {\bf 91} (2003) 232002.
\bibitem{ZhuWang} P. Z. Huang, W. Z. Deng, X. L. Chen and S. L. Zhu,
Phys. Rev. {\bf D69} (2004) 074004;  Z. G. Wang, W. M. Yang and S.
L. Wan, hep-ph/0501278.

\bibitem{Ioffe} B. L. Ioffe, Nucl. Phys. {\bf B188} (1981) 317, Erratum-ibid. {bf B191} (1981) 591.

\bibitem{Reducible} Y. Kondo, O. Morimatsu, T. Nishikawa, Phys. Lett. {\bf B611} (2005)
93;  S. H. Lee, H. Kim, Y. Kwon, Phys. Lett. {\bf B609} (2005) 252;
Y. Kwon, A. Hosaka, S. H. Lee, hep-ph/0505040.

\bibitem{WaveAdd} C. E. Carlson, C.
D. Carone, H. J. Kwee, V. Nazaryan, Phys. Rev. {\bf D70} (2004)
037501; F. Buccella, P. Sorba, Mod. Phys. Lett. {\bf A19} (2004)
1547.

\bibitem{JW03} R. L. Jaffe, F. Wilczek, Phys. Rev. Lett. {\bf 91} (2003) 232003.
\bibitem{KL}  M. Karliner, H. J. Lipkin, hep-ph/0307243;  M. Karliner,
 H. J. Lipkin, Phys. Lett. {\bf B575} (2003) 249.



 \bibitem{JM04} B. K. Jennings, K. Maltman, Phys. Rev. {\bf D69} (2004)
 094020;  J. J. Dudek, F.E. Close, Phys. Lett. {\bf  B583} (2004)
 278; D. Melikhov, S. Simula, B. Stech, Phys. Lett. {\bf B594} (2004)
 265.

\bibitem{color-spin-flavor} C. E. Carlson, C. D. Carone, H. J. Kwee, V.
Nazaryan, Phys. Lett. {\bf B573} (2003) 101; F. Stancu, D. O. Riska,
Phys. Lett. {\bf B575} (2003) 242; C. E. Carlson, C. D. Carone, H.
J. Kwee, V. Nazaryan, Phys. Lett. {\bf B579} (2004) 52; R. Bijker,
M. M. Giannini, E. Santopinto, Eur. Phys. J. {\bf A22} (2004) 319.


\bibitem{Ioffe84}  B. L. Ioffe and A. Smilga,  Nucl. Phys. {\bf B232}  (1984)
109.
\bibitem{Ioffe82}  V. M. Belyaev, B. L. Ioffe, Sov. Phys. JETP {\bf 56} (1982)
493.



\end{thebibliography}
\end{document}